\documentclass[lettersize,journal]{IEEEtran}
\usepackage{amsmath,amsfonts}
\usepackage{algorithmic}
\usepackage{algorithm}
\usepackage{array}
\usepackage{todonotes}
\usepackage{textcomp}
\usepackage{stfloats}
\usepackage{url}
\usepackage{verbatim}
\usepackage{graphicx}
\usepackage{cite}
\usepackage{xcolor}
\usepackage{tipa}
\usepackage{multirow}
\usepackage{hyperref}
\usepackage{booktabs}

\definecolor{arytenoid-cartilage}{rgb}{0.54, 0.17, 0.89} % blueviolet
\definecolor{epiglottis}{rgb}{0.25, 0.88, 0.82} % turquoise OK
\definecolor{lower-lip}{rgb}{0.00, 1.00, 0.00} % lime OK
\definecolor{pharyngeal-wall}{rgb}{0.85, 0.66, 0.13} % goldenrod 
\definecolor{soft-palate-midline}{rgb}{0.12, 0.56, 1.00} % dodgerblue
\definecolor{tongue}{rgb}{1.00, 0.55, 0.00} % darkorange
\definecolor{upper-lip}{rgb}{1.00, 0.00, 1.00} % magenta
\definecolor{vocal-folds}{rgb}{1.00, 0.41, 0.71} % hotpink
\definecolor{projected-points}{rgb}{0.0, 0.0, 0.545} % darkblue
\definecolor{upper-incisor}{rgb}{1,1,0}

\definecolor{HeightVT}{RGB}{0,128,0}          % green
\definecolor{LA}{RGB}{123,104,238}            % mediumslateblue
\definecolor{LD}{RGB}{85,107,47}              % darkolivegreen
\definecolor{TTCD}{RGB}{139,69,19}            % saddlebrown
\definecolor{TBCD}{RGB}{189,183,107}          % darkkhaki
\definecolor{TRCD}{RGB}{152,251,152}          % palegreen
\definecolor{LH}{RGB}{220,20,60}          % crimson
\definecolor{LP}{RGB}{216,191,216}             % thistle
\definecolor{TRCL}{RGB}{238,130,238}

\newif\ifshowtodos
\begin{document}
\title{Acoustic-to-articulatory Inversion of the Complete Vocal Tract from RT-MRI with Various Audio Embeddings and Dataset Sizes}
%2 eme titre : Investigating Audio Embeddings and Dataset Size for Full Vocal-Tract Inversion from RT-MRI

\author{%
Sofiane Azzouz,
Pierre-André Vuissoz,
Yves Laprie
%}
\thanks{Sofiane Azzouz and Yves Laprie are with the Université de Lorraine, CNRS, Inria, F-54000 Nancy, France (e-mail : sofiane.azzouz@loria.fr, yves.laprie@loria.fr).}
\thanks{Pierre-André Vuissoz is with the Université de Lorraine, Inserm, IADI U1254, F-54000 Nancy, France (e-mail : pa.vuissoz@chru-nancy.fr).}}
\maketitle

\begin{abstract}
%Articulatory-to-acoustic inversion strongly depends on the type of data used. While most previous studies rely on EMA, which is limited by the number of sensors and restricted to accessible articulators, we propose an approach aiming at a complete inversion of the vocal tract, from the glottis to the lips. To this end, we used approximately 3.5 hours of RT-MRI data from a single speaker and automatically segmented the articulatory contours. The audio was also denoised, and the model was based on a Bi-LSTM architecture.
Articulatory-to-acoustic inversion strongly depends on the type of data used. While most previous studies rely on EMA, which is limited by the number of sensors and restricted to accessible articulators, we propose an approach aiming at a complete inversion of the vocal tract, from the glottis to the lips. To this end, we used approximately 3.5 hours of RT-MRI data from a single speaker.
The innovation of our approach lies in the use of articulator contours automatically extracted from MRI images, rather than relying on the raw images themselves. By focusing on these contours, the model prioritizes the essential geometric dynamics of the vocal tract while discarding redundant pixel-level information. These contours, alongside denoised audio, were then processed using a Bi-LSTM architecture.
Two experiments were conducted: (1) the analysis of the impact of the audio embedding, for which three types of embeddings were evaluated as input to the model (MFCCs, LCCs, and HuBERT), and (2) the study of the influence of the dataset size, which we varied from 10 minutes to 3.5 hours.
Evaluation was performed on the test data using RMSE, median error, as well as Tract Variables, to which we added an additional measurement: the larynx height. The average RMSE obtained is 1.48\,mm, compared with the pixel size (1.62\,mm). These results confirm the feasibility of a complete vocal-tract inversion using RT-MRI data.

\end{abstract}
% Doit etre 150 et 250 mots.


%PAV
\ifshowtodos
    \todo[inline]{PAV: Just comment "/ showtodostrue"  }
\fi
%PAV

\section{Introduction}
Articulatory-to-acoustic inversion consists in recovering the shape of the vocal tract from the speech signal. Early work on inversion was based on an analysis-by-synthesis paradigm. One of the first examples is Wakita's inverse filtering approach \cite{wakita1973}, which imposes a highly constrained model in the form of an all-pole model to enable inversion, with the advantage of limiting the use of articulatory data. Indeed, Wakita points out that it is difficult to acquire data, that this poses risks for subjects (as the data was obtained through X-rays), and that the techniques available at the time did not allow for the transverse dimension (perpendicular to the mid-sagittal plane) to be determined.
The entire history of articulatory acoustic inversion is a kind of compromise between the complexity of an analysis model and the existence of data to adjust or replace the model. Following Wakita, the analytical approach was improved in terms of both acoustic numerical simulations and the geometric model of the vocal tract.

Recent modeling techniques are based on acoustic simulations that take into account vocal folds, aerodynamics, and the acoustic properties of the vocal tract. They aim to achieve a formulation that is sufficiently efficient in terms of computation time and fidelity to physics. Advances in acoustic simulations now make it possible to produce high-quality speech, as demonstrated by the VocalTractLab system \cite{Birkholz2013}. The shape of the vocal tract and its temporal evolution must be specified as input for these simulations. The first technique uses geometric primitives, cylinders whose center and radius can change, and planes in the case of a three-dimensional model \cite{Birkholz2003}. This technique has the advantage of being simple, but it does not guarantee the realism of the geometric shape. A second technique consists of using medical images of the vocal tract (X-ray images in the 1970s and, since then, 2D MRI images or 3D MRI volumes), which guarantees greater realism but presents the difficulty of switching from one speaker to another. Recently, Gao and Birkholz \cite{GaoYingmingBirkholz2024} have used Birkholz's articulatory synthesizer to propose an articulatory copy synthesis system, the principle of which is to search for the parameters of the articulatory synthesis model that generate a speech signal close to  the input speech. The principle is therefore very similar to articulatory acoustic inversion, and the resynthesized speech appears to be of very good quality. Despite these advances, there is still a discrepancy between the underlying geometric analysis model and the actual speaker, which compromises the geometric accuracy of the inversion. 

The second approach involves dispensing with an analysis model and establishing a direct link between the signal and the geometry of the vocal tract. With the advent of new techniques for acquiring articulatory data, first microbeam X-ray technology (abandoned due to the risks posed by X-rays), and especially EMA (ElectroMagnetic Articulography), it has become possible to acquire a sufficient volume of data to perform machine learning \cite{Wrench2000, Richmond2011}. Unfortunately, this data only covers a few points (3 or 4 on the tongue, 1 on the soft palate, 1 on the lower central incisor, and 2 on the lips), but it is possible to record speech for a little over 30 minutes. This technique has become widely used because, unlike the microbeam technique, it does not pose any radiological risks. On the other hand, the fact that the sensors are connected to the measuring device by wires slightly alters speech articulation, and these sensors often come off after about 30 minutes. The second weakness is that the articulatory information is very limited, since only the positions of seven points are known, and these points are located at the front part of the vocal tract, providing no information about the pharynx and larynx. Finally, if several acquisition sessions are made, it is impossible to guarantee that the sensors will be glued in the same place. 
In practice, evaluations generally focus on distances between two sensors, or between a sensor and a reference contour such as the palate.

This data has been used for training stochastic and neural approaches. The weakness of stochastic techniques is that temporal modeling, which is essential for speech, is insufficient. This explains why neural network based approaches quickly supplanted stochastic models. The architecture of neural networks has evolved from multilayer networks \cite{Richmond2006} to recurrent networks, LSTMs or Bi-directional Gated Recurrent Units \cite{parrot2020, Illa2020, Mcghee2024} which have become the standard in recent years.

However, these studies only provide partial information about the vocal tract, as they only concern the anterior part of the oral cavity, whereas it is well known that the length of the vocal tract —and therefore the position of the larynx— has a strong influence on resonance frequencies, and that there are compensation mechanisms involving the lower part of the vocal tract. Despite their interest in terms of taking the temporal dimension into account, the use of inversion trained on EMA data is not realistic, because, from an application point of view, it is essential that the entire vocal tract be covered.

Only dynamic MRI data can meet this requirement, but this necessitates high-quality data to provide reliable geometric information in sufficient quantity \cite{Lim2021}, good spatial resolution images and high-quality denoised speech signal despite the strong noise generated by the MRI machine \cite{Isaieva2021, Maekawa2025}. Real-time Magnetic Resonance Imaging (rt-MRI) was introduced as an alternative to EMA, with the first recorded corpus described in \cite{ramanarayanan2018analysis}. Deep learning approaches have been applied to these data \cite{Csapo2020}. However, the use of rt-MRI remains limited due to several constraints: difficulty in acquiring sufficiently large datasets, low signal quality after denoising, lack of robust contour-tracking tools, and relatively low spatial resolution (68x68 pixels with a voxel size of 2.9\,x\,2.9\,x\,5mm$^3$ or more recently 84x84 pixels with a voxel size of 2.4\,×\,2.4\,x\,6 mm$^3$ in~\cite{Lim2021}), along with certain MRI-related artefacts.

%PAV
\ifshowtodos
    \todo[inline]{PAV: Ce serait bien si tu arrive aussi a donner la taille du voxel, quelque chose comme $2 \times 2 \times 8 mm^3$ à vérifier les dimention exacte dans la publication originelle.  }
\fi
%PAV

There are two ways to perform inversion. The first is implicit inversion, which involves recovering an MRI image. \cite{Csapo2020, Oura2024} use an LSTM-based approach to recover the image, while \cite{Nguyen2024} uses a diffusion model used in image synthesis. In both cases, the evaluation is performed on the image pixels, either in terms of distance or in terms of global or local correlation. In \cite{Nguyen2024}  correlation focuses on a few regions of interest to the vocal tract and not the entire image. 

In all cases, the evaluation does not take into account any information about the position of the articulators themselves. The first weakness is that the assessment therefore does not take into account any information relating to articulators.
The second weakness is that exploiting the inversion results  requires a post-processing step. For example, to use the inversion as input for articulatory synthesis, the contours of the articulators must be extracted from the images reconstructed during the inversion. These images are marred by inversion errors and are therefore less easy to interpret and segment automatically. 

The second avenue of research consists of working on the contours of the articulators, which must therefore be extracted automatically from the images. The advantage is that the inversion result is directly usable since the inversion provides the contours of the articulators from the signal. Our work has shown that it is possible to extract the contours automatically with a high degree of reliability \cite{ribeiro:hal-04376938}. We chose the second approach, which has the advantage of providing information that is easy to use as articulatory feedback in educational applications or speech rehabilitation, for example.

%PAV
\ifshowtodos
    \todo[inline]{PAV: "easy to use in applications." le lecteur ne peut pas voir de quel application tu parle et en quoi ça les facilite. Donne un exemple où le fait d'avoir le contours directement élimine une étape de resegementation ou avoir les countours permetterait autre chose mais quoi ? Et en quoi une étape de segementation supplémentaire est vraiment un problème tel que c'est pas "easy to use" ??  }
\fi

In our study, we used an rt-MRI dataset with good spatial resolution (136×136 pixels), good quality denoised speech signal of higher quality than most existing rt-MRI datasets.

%PAV
\ifshowtodos
    \todo[inline]{PAV: Idem $1.41 \times 1.41 \times 8 mm^3$ . Désolé ça c'est la dimention pour les dernières donnée acquises avec le nouveau protocole qui a changé avec la mpi90 alors qu'en mpi62 on acquière bien en $1.62 \times 1.62 \times 8 mm^3$  }
\fi
%PAV

%Furthermore, instead of inverting the complete rt-MRI images, we opted for automatic contour tracking to extract the shapes of individual articulators, and performed inversion only on these contours rather than on the full image. We use this high-quality database for both image and sound to assess the geometrical accuracy of inversion.

Furthermore, in contrast to previous studies that invert complete RT-MRI images, we opted for automatic contour tracking to extract the shapes of individual articulators, and performed inversion only on these contours rather than on the full image. We use this high-quality database for both image and sound to assess the geometrical accuracy of inversion.

%PAV
\ifshowtodos
    \todo[inline]{PAV: "real potential of inversion" mais encore on a pas quelque chose de concret a évaluer. Qu'entend t'on par "real potoential" ? }
\fi
%PAV
Until now, the articulatory variables used to characterize the vocal tract from a phonetic point of view have not taken into account the vertical position of the larynx, which plays an essential role in modifying the length of the vocal tract and, consequently, its resonance frequencies. The reason for this is that it is impossible to reliably locate the position of the larynx using EMA data. As we use high-quality MRI data and contours, we have added this new articulatory variable to our work.

We also use a large single-speaker database, which allows us to study in detail various aspects of training, such as the impact of dataset size on inversion performance — using subsets of 10 minutes, 30 minutes, 1 hour, 2 hours, and the full dataset.
In this paper, we propose a complete vocal tract inversion approach from the glottis to the lips, striving to work under optimal conditions on a single speaker, in order to achieve the lowest possible error rate for MRI data. Furthermore, we precisely determine the minimal amount of data required to obtain reliable and reproducible results for one speaker.

%PAV
\ifshowtodos
    \todo[inline]{PAV: Il me semble que dans l'introduction vous n'avez pas montré que l'analyse sur la quantité de donnée nécessaire n'existe pas et est importante a réalisé. Et accentué la quantité de donnée que les autres ont utilisé ? }
\fi
%PAV

%PAV
\ifshowtodos
    \todo[inline]{PAV: Le dernier paragraphe manque de concision, il faut vraiment accentuer ce que l'on démontre dans le papier, qu'est ce que l'on prouve dans nos résultat.}
\fi
%PAV

\section{Dataset}
    \subsection{corpus}
    The corpus was recorded at the Centre Hospitalier Régional de Nancy. It consists of recordings from a single female French speaker and contains 2,100 sentences, corresponding to approximately 3.5 hours of speech. The dataset is structured into 153 acquisitions, each lasting 80 seconds and containing 4,000 images acquired at a rate of 50 images per second. Each image corresponds to a 8 mm slice in the mid-sagittal plane and has a resolution of 136×136 pixels and a pixel spacing of 1.62 mm. The corresponding audio was recorded using an optical microphone at 16 kHz and then  denoised with the algorithm proposed in \cite{Ozerov2012}. The audio quality of denoised speech is close to clean speech (an example audio file is provided in the Supplementary Material).
    %\textcolor{red} {il faudra un lien sur un exemple ou inclure un fichier}
    In addition, we also have precise phonetic segmentations. The ASTALI software \cite{fohr2015importance} was used to perform a forced alignment, which was then carefully manually corrected by an expert. The corrections focused on the boundaries between sounds, most often for plosives. In addition, we separated the closure of voiceless plosives from the burst because the articulatory position changes significantly between the two parts. We did not apply this correction to voiced plosives because it is often more difficult to distinguish between the two parts.
    
\iffalse
%PAV
\ifshowtodos
    \todo[inline]{PAV: Désolé mais un article doit être reporductible, ici il faut une courte description de ce qui a été corrigé.}
\fi
%PAV
\fi

    \subsection{Image registration}
\begin{figure}[ht]
    \centering
    \includegraphics[width=0.3\textwidth]{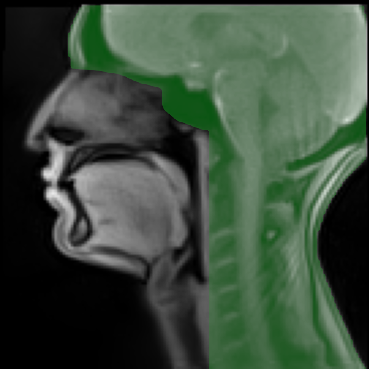}
    \caption{mask of the reference image. The green region represents the static part of the vocal tract}
    \label{fig:mask}
\end{figure}

    Our data was recorded over six different sessions. In order to ensure that the speaker maintains the same posture we made a blocking foam perfectly adapted to the MRI antenna and the speaker’s head. Despite these precautions, it is impossible for the speaker to maintain exactly the same posture during the six recording sessions and even during the same session, which could lead to misaligned images.  
    To overcome this difficulty, we realigned the images for each acquisition. A mask covering only the static part of the head (see~\autoref{fig:mask}) was created, after which rigid transformations were applied within a ±4-pixel translation range and ±4° rotation range. The transformation yielding the highest normalized cross-correlation with a reference image was then selected.

\iffalse       
%PAV
\ifshowtodos
    \todo[inline]{PAV: Les 4 pixels c'est pour la translation mais combien de degré pour la rotation ? }
\fi
%PAV
\fi

\subsection{Choice of input representation}

    We explored several types of speech signal modeling at the input of inversion. Traditionally, MFCCs are used for inversion because they have long provided the best results for automatic speech recognition \cite{9955539}.
    One of the strengths of MFCCs is that they neutralize the influence of the fundamental frequency and attenuate differences between speakers by using a non-linear frequency scale. Here, we use data from a single speaker and we thus also used linear cepstral coefficients (LCCs) \cite{rabiner1978digital}, which use a linear frequency scale. This should be an advantage because the formants that depend directly on the vocal tract shape are preserved. 
    
    Self-supervised learning (SSL) models have recently demonstrated their effectiveness in various speech processing tasks, particularly in articulatory acoustic inversion. In \cite{cho2023evidence}, several SSL models — such as HuBERT~\cite{hsu2021hubert}, Wav2Vec2~\cite{baevski2020wav2vec}, and WavLM~\cite{chen2022wavlm} — were compared to traditional approaches based on hand-crafted acoustic features, such as Mel-frequency cepstral coefficients (MFCCs), for the inversion task on EMA data. The results showed that HuBERT outperformed all other methods. Similarly, \cite{attia2024improving} conducted experiments on XRMB data and compared the representations extracted by HuBERT to those based on MFCCs, confirming HuBERT’s superiority through higher correlations between acoustic and articulatory signals. 
%PAV
\ifshowtodos
    \todo[inline]{PAV: Je me demande si la digression ci dessus sur la représentation sonore idéale pour l'inversion n'est pas a mettre dans l'introduction ? C'est pas vraiment des methodes à porprement parlé mais des résultats historique que tu mentionne ici. }
\fi
%PAV
In this work we will therefore only compare the results obtained with MFCCs, LCCs, and HuBERT\footnote{\url{https://huggingface.co/docs/transformers/model_doc/hubert}}.

\subsection{Data preprocessing}
\begin{figure}[b] % Utiliser l'environnement figure
  \centering
  \includegraphics[width=0.475\textwidth]{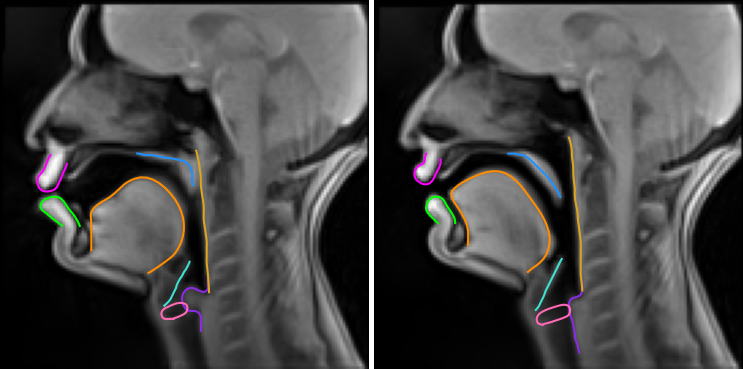} % Spécifier le chemin et la taille
  \caption{
    Segmentation of articulators contour tracked in two images of the rt-MRI film:\hspace{0.2cm}
    \raisebox{0.25ex}{\textcolor{arytenoid-cartilage}{\rule{0.2cm}{0.08cm}}} Arytenoid cartilage,
    \raisebox{0.25ex}{\textcolor{epiglottis}{\rule{0.2cm}{0.08cm}}} Epiglottis,
    \raisebox{0.25ex}{\textcolor{lower-lip}{\rule{0.2cm}{0.08cm}}} Lower lip,%\hspace{-0.1cm}
    \raisebox{0.25ex}{\textcolor{vocal-folds}{\rule{0.2cm}{0.08cm}}} Vocal folds,\hspace{0.135cm}
    \raisebox{0.25ex}{\textcolor{soft-palate-midline}{\rule{0.2cm}{0.08cm}}} Soft palate midline,\hspace{0.05cm}
    \raisebox{0.25ex}{\textcolor{tongue}{\rule{0.2cm}{0.08cm}}} Tongue,%\hspace{0.17cm}
    \raisebox{0.25ex}{\textcolor{upper-lip}{\rule{0.2cm}{0.08cm}}} Upper lip,%\hspace{0.05cm}
    \raisebox{0.25ex}{\textcolor{pharyngeal-wall}{\rule{0.2cm}{0.08cm}}} Pharyngeal wall
  }
  \label{fig:original_contours} % Ajouter un label pour les références
\end{figure}

Unlike other works using full MRI images, the proposed method recovers the contours of articulators. The vocal tract is represented by the contours of eight articulators or cartilages: the upper lip, lower lip, tongue, soft palate midline (velum), pharyngeal wall, epiglottis, arytenoid cartilage, and vocal folds (glottis). Additionally, two landmarks are included that do not contribute directly to the inversion—the lower incisor and upper incisor. However, the upper incisor is utilized in the calculation of the Tract Variables (TVs), inspired by articulatory phonology~\cite{browman1992articulatory}. All contours, as shown in~\autoref{fig:original_contours}, were obtained using an automatic tracking approach based on RCNN~\cite{Ribeiro2021}, developed by our team. This approach, which maintains a tracking error of approximately 1.00 mm, is available as open-source code\footnote{\url{https://github.com/vribeiro1/vocal-tract-seg}}.

Each articulator contour consists of 50 points with X and Y coordinates.
The contours were normalized following the approach used in \cite{parrot2020}. For each contour point a moving average was computed over the 30 preceding and following frames. This average was  subtracted from the coordinates, and the result divided by the corresponding standard deviation for each session.

Three different representations were used as input features: MFCCs, LCCs, and HuBERT embeddings.
The MFCCs were computed together with their first-order (\(\Delta\)) and second-order (\(\Delta\Delta\)) derivatives, using 13 coefficients, a window size of 25 ms, and a hop length of 10 ms.
The LCCs were extracted using the first 30 linear cepstral coefficients with the same window and hop sizes.
%For HuBERT, we compared both the Base and Large models, and the best results were obtained with the Base model. We also are concsious taht HuBERT was trained on english multi-speaker this we also see in the litterarautre that there is a multi language fine tuned mHuBERT-147 who have the same reults. Therefore, in this study, only the HuBERT-Base representation was used. HuBERT processes 16 kHz audio and outputs embeddings of dimension 768 at a frame rate of 50 Hz.
For HuBERT, we compared both the Base and Large architectures; the best results were obtained using the Base model. While the standard HuBERT was trained on English multi-speaker data, the literature indicates that the multilingual fine-tuned version, mHuBERT-147 \cite{boito2024mhubert}, yields very similar results, what our preliminary tests have confirmed for inversion. Therefore, this study utilizes only the HuBERT-Base representation. HuBERT processes 16 kHz audio and generates 768-dimensional embeddings at a frame rate of 50 Hz.
All features (MFCCs, LCCs, and HuBERT) were normalized per session by subtracting their mean and dividing by their standard deviation.

\section{Methods}
    \subsection{Model architecture}
\begin{figure*}[b]
        \centering
        \includegraphics[width=1\textwidth]{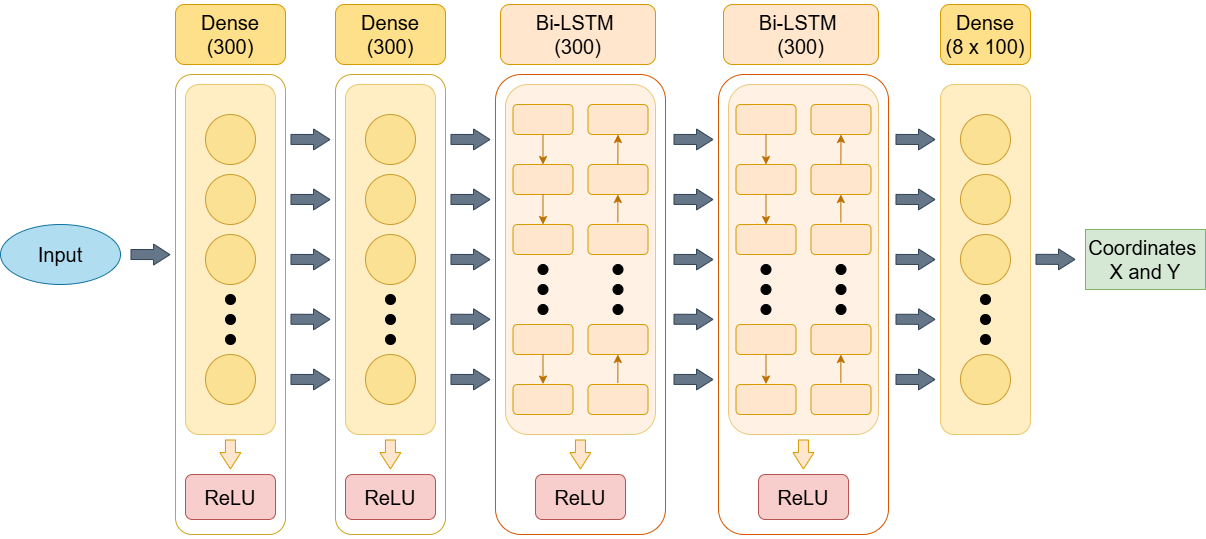} % Remplacer par l'image
        \caption{Architecture of the model}
        \label{fig:model_architecture}
    \end{figure*}

    Our model is based on a bidirectional LSTM (Bi-LSTM) architecture~\cite{azzouz2025complete, azzouz2025reconstruction}, which has demonstrated strong performance for this specific task. The choice of a Bi-LSTM is motivated by its inherent ability to capture bidirectional temporal dependencies, which are crucial for modeling the anticipatory and carryover coarticulation effects~\cite{farnetani1997coarticulation} in speech production. The architecture consists of five layers and takes acoustic feature vectors as input, passes through two dense layers with 300 units each, followed by two bidirectional LSTM layers, each consisting of 300 units (see~\autoref{fig:model_architecture}).
    
    The output is generated by a dense layer, producing a tensor of size 100 × 8 (number of articulators), where 100 represents the contour coordinates (50 for the X coordinates and 50 for the Y coordinates). In our previous work \cite{azzouz2025reconstruction}, we tested several variants of this model, including an 8-layer version, and we concluded that this model is sufficient to perform this task.
    
    \subsection{Loss function}
    Since acoustic-to-articulatory inversion is a regression task, the Mean Squared Error (MSE) is the most commonly used loss function. \begin{equation}
    \text{MSE} = \frac{1}{n} \sum_{i=1}^{n} \left( y_i - \hat{y}_i \right)^2\label{eq1}
    \end{equation}
    where n represents the total number of the contour points, \(y_i\) and \(\hat{y}_i\) represent the true and predicted values of the output for example \(i\), respectively.
    
    We use the sum of the Mean Squared Error for all the contours of each frame to minimize the distance between the predicted and the ground-truth articulatory positions.

    \subsection{Evaluation}
    First, the data were denormalized by multiplying with the standard deviation and adding the mean used to normalize each articulator. we evaluated the model using the average RMSE. For each image, we computed the mean RMSE for each articulator, by considering its 100 coordinates. Then, we calculated the mean error across all articulators, which gives us the image-level RMSE. After obtaining the RMSE for each image, we computed the median as well as the overall mean RMSE across all images. These values are expressed in millimeters (mm).
    \begin{equation}
        \text{RMSE} = \sqrt{\frac{1}{n} \sum_{i=1}^{n} \left( y_i - \hat{y}_i \right)^2}\label{eq2}
    \end{equation}

    \section{Articulatory trajectories}
\begin{figure}[b]
        \centering
        \includegraphics[width=0.3\textwidth]{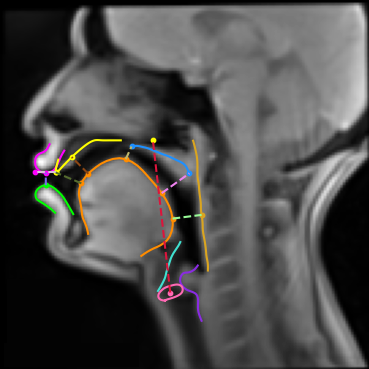} % Remplacer par l'image
          \caption{
    Visual Representation of the Various Vocal Tract Variables (TVs) represented by dashed lines :\hspace{0.2cm}
    \raisebox{0.25ex}{\textcolor{LP}{\rule{0.2cm}{0.08cm}}} LP,
    \raisebox{0.25ex}{\textcolor{LA}{\rule{0.2cm}{0.08cm}}} LA,
    \raisebox{0.25ex}{\textcolor{LD}{\rule{0.2cm}{0.08cm}}} LD, %\hspace{-0.1cm}
    \raisebox{0.25ex}{\textcolor{TTCD}{\rule{0.2cm}{0.08cm}}} TTCD,
    \raisebox{0.25ex}{\textcolor{TBCD}{\rule{0.2cm}{0.08cm}}} TBCD, \\
    \raisebox{0.25ex}{\textcolor{TRCL}{\rule{0.2cm}{0.08cm}}} TRCL,%\hspace{0.05cm}
    \raisebox{0.25ex}{\textcolor{TRCD}{\rule{0.2cm}{0.08cm}}} TRCD,%\hspace{0.17cm}
    \raisebox{0.25ex}{\textcolor{LH}{\rule{0.2cm}{0.08cm}}} LH
  }
        \label{fig:Articulatory_variable}
    \end{figure}

    The acoustic impact of a geometric error depends on the distance between the two walls of the vocal tract, usually a fixed wall, such as the palate, and a mobile articulator, such as the tongue. The smaller this distance, and therefore the greater the constriction, the greater the acoustic impact, which determines the realization of a phonetic feature. An articulator is critical for a phoneme if it corresponds to a minimum distance, i.e. a constriction, and the tract variables (VT) of articulatory phonology \cite{browman1992articulatory} correspond to the points or regions of the vocal tract that play a critical phonetic role for one or more phonemes. Unlike global metrics such as RMSE applied to articulatory contours, TVs thus allow for targeted evaluation of the articulatory gestures specific to each phoneme. These distances provide a more functionally meaningful assessment of articulatory inversion quality, as they reflect constrictions formed in the vocal tract and they are directly linked to constriction regions in the vocal tract. They were defined within the framework of gestural phonology, shown in the~\autoref{fig:Articulatory_variable}, and the tract variables with their corresponding constrictors are presented in \autoref{tab:articulatory_variables}.

    Unlike EMA data, we do not have fixed points directly located on the critical articulators required for computing the TVs. In our case, for each TV, we selected a set of points on the two articulators in order to approximate as closely as possible the anatomical reference points.

    We have added the larynx height (see~\autoref{fig:Articulatory_variable} and~\autoref{larynx}) as a new variable that we believe to be important. %The first is the distance between the uvula and the back of the tongue, which is minimal for \textipa{/\textinvscr/} in French.
    It does not correspond to a constriction, but has a significant acoustic impact on resonance frequencies. 

\iffalse
%PAV
\ifshowtodos
    \todo[inline]{PAV: La table I donne une définition cannonique des "Vocal tract variables", mais c'est pas la définition que tu applique programatiquement, tu l'explique bien pour VEL dans le paragraphe A Velum movement, et pour LH dans le paragraphe B. Larynx height mais tu doits aussi le faire pour les autres variable peut être plus succintement et en annexe mais ça manque. Example LA : $norm(min(y_{upperlip-contour}) - max(y_{lowerlip-contour}))$ ou quelque chose d'autre suivant ton implémentation. C'est important d'avoir les formule exacte.
    Yves : ces variables articulatoires, sauf celle pour la position du larynx, sont "massivement utilisées" et ce n'est pas la peine de redonner leur définition. Ici, ça peut changer un peu leur calcul qui ne repose pas sur la position d'un seul capteur. La figue 4 me semble assez explicite. à discuter
    }
\fi
%PAV
\fi

    \begin{table*}[hb]
    \caption{Vocal tract variables and their associated constrictor distances.}
    \label{tab:articulatory_variables}
    \centering

    % --- ZONE DE RÉGLAGES MANUELS ---
    %\footnotesize      % Tu peux changer par \small ou \scriptsize selon tes besoins
    \small
    % \setlength{\tabcolsep}{12pt}    % Augmente pour élargir le tableau
    % \renewcommand{\arraystretch}{1.3} % Augmente pour aérer verticalement
    % -------------------------------

    \begin{tabular}{lll} % Utilisation de 'l' partout pour un alignement propre à gauche
        \toprule
        & \textbf{VT Variable} & \textbf{Definition / Distance} \\
        \midrule
        \textbf{LA}   & Lip Aperture                    & Vertical distance between the upper and lower lips (mouth opening) \\
        \textbf{LP}   & Lip Protrusion                 & Forward displacement of the lips along the anterior-posterior axis \\
        \textbf{LD}   & Lip Constriction Degree        & Distance between the lower lip and the upper teeth \\
        \textbf{TTCD} & Tongue Tip–Palate Distance     & Distance between the tongue tip and the hard palate \\
        \textbf{TBCD} & Tongue Body–Palate Distance    & Distance between the tongue body and the hard palate \\
        \textbf{TRCD} & Tongue Root Constriction Degree & Distance between the tongue root and the posterior pharyngeal wall \\
        \textbf{TRCL} & Tongue Root Constriction Location & Position along the vocal tract where the tongue root occurs \\
        \textbf{VEL}  & Velic Opening Distance         & Distance between the velum and the posterior pharyngeal wall \\
        \textbf{LH}   & Larynx height                  & Vertical distance between the glottis and the hard palate \\
        \bottomrule
    \end{tabular}
\end{table*}

    \subsection{Velum movement}
    The velum is elongated in shape, and we chose to use its midline because it offers greater robustness in terms of tracking and modeling. The main movement of the velum corresponds to an opening/closing that allows or prevents airflow into the nasal cavities. An initial idea is to measure the distance between the lower end of the midline, i.e. the velum tip, and the pharyngeal wall, but this has two weaknesses. The first is that the distance between this point and the pharyngeal wall is never zero since it is in the middle of the velum and therefore does not contact this wall. The second weakness is that the velum thickness is minimal at this point, which reduces the image contrast and consequently reduces the robustness of the tracking.

\begin{figure}[t]
        \centering
        \includegraphics[width=0.475\textwidth]{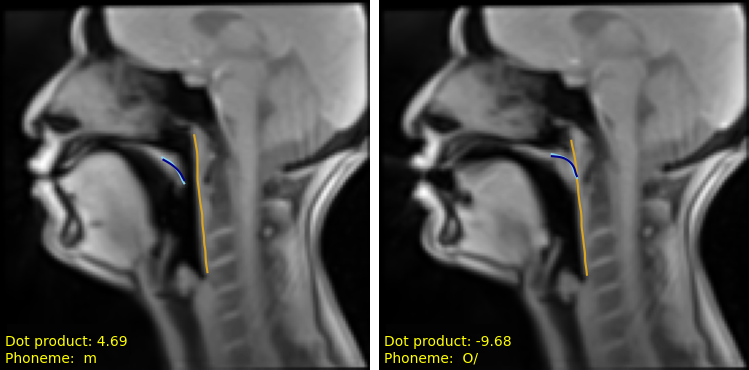} % Remplacer par l'image
        \caption{Example of two velopharyngeal port opening configurations:\\open (4.69) and closed (–9.68).
        %Velopharyngeal port opening score for three images (from left to right, i.e. from an open position to a closed position): 4.69 , 0.13, and -9.68.\\
        \raisebox{0.25ex}{\textcolor{soft-palate-midline}{\rule{0.2cm}{0.08cm}}} Original points,\hspace{0.05cm} 
        \raisebox{0.25ex}{\textcolor{projected-points}{\rule{0.2cm}{0.08cm}}} Projected points.\hspace{0.05cm}
        }

        \label{fig:pca_projection}
    \end{figure}

    To estimate the dynamics of velum opening and closing we thus  characterize velopharyngeal movement more comprehensively by defining a subset of 25 points along the midline, starting at the velum tip and extending posteriorly (see~\autoref{fig:pca_projection}). This region was chosen because it corresponds anatomically to the portion of the velum most directly involved in the velopharyngeal closure gesture, thereby providing a detailed representation of its motion during both lowering (opening) and raising (closing) phases.
    A principal component analysis (PCA) was applied to these 25 midline points across the entire dataset including all silences, with all articulatory coordinates centered and normalized prior to analysis to ensure that differences in absolute position do not bias the results.  

\iffalse
\ifshowtodos
\todo[inline]{ Scaled est risqué car on peut penser qu'il y a un effet de d'échelle ce qui n'est pas le cas. Ce qui répond peut-être à la question de Pierre-André}
\fi
\fi

\iffalse
%PAV
\ifshowtodos
    \todo[inline]{PAV: "coordinates centered and scaled" : C'est peut être trop long de s'étendre ici mais tu devrais donné quelque part, peut être en annexe comment est opérer le centrage et la mise a l'échelle, surtout que tu a différente coordonée entre les coordonnée détectée par l'algo de Vinicius et celle normalisée en entrée de ton modèle, c'est pas les même et le centrage et la mise a l'échelle n'est pas formément la même opération et de plus ça implique aussi une moyenne et sur quel ensemble sujet/sequence/glissante ? }
\fi
%PAV
\fi
    Similarly to \cite{cunha2024physiological} the first principal component PC1, which accounted for 57\% of the total variance in our dataset (59\% in \cite{cunha2024physiological}), was interpreted as representing the primary degree of freedom corresponding to the opening/closing gesture of the velum.

    %PAV
\iffalse
\ifshowtodos
    \todo[inline]{PAV: "representing the canonical opening/closing gesture of the velum" : Même is c'est trivial il n'est pas évident que ce soit toujours la première composante, ca doit faire partie de tes résultats, tu fait la décomposition et tu observe que la PC1 correspond bien à la fermeture/ouverture si tu a la place met juste une image dans les résultat avec trois position le long de la PC1 (max, median, min) pour bien montrer l'ouverture fermeture et le commentaire ci dessous tu le met en disscussion.  Yves : J'ai fait référence à l'article Conceição Cunha, Phil Hoole... ce qui permet de répondre à la remarque de Pierre-André.}
\fi
\fi
%PAV
For each frame in the dataset, the degree of velum opening or closure was quantified by projecting the 25-point midline configuration onto PC1. This projection, i.e. the dot product between the PC1 loading vector and the current velum configuration, yielded a scalar score that we term the PC1 score. The polarity of PC1 was defined such that positive scores indicate motion in the opening direction (velopharyngeal port enlarging), whereas negative scores indicate motion in the closing direction (velopharyngeal port constricting). Thus, in the context of our analysis, large positive PC1 scores correspond to maximal opening, large negative scores to maximal closure, and values near zero to intermediate or neutral positions (see~\autoref{fig:pca_projection}).

    By reducing the velopharyngeal movement to this single physiologically interpretable dimension, we were able to track the temporal evolution of the gesture with high precision.

    \subsection{Larynx height} \label{larynx}
    
    In addition to the vocal tract variables (TVs) already studied in the literature \cite{browman1992articulatory}, we introduced a new metric that consists of measuring the larynx height.

\begin{figure}[ht]
    \centering
    \includegraphics[width=0.3\textwidth]{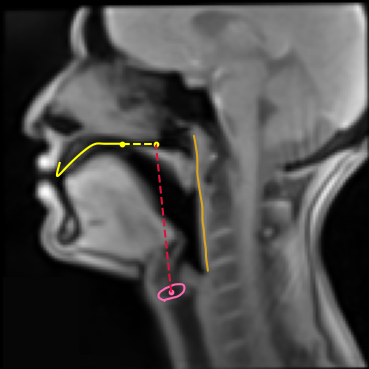}
    \caption{Visual Representation of the Larynx height \\
        \raisebox{0.25ex}{\textcolor{LH}{\rule{0.2cm}{0.08cm}}} LH
        \raisebox{0.25ex}{\textcolor{upper-incisor}{\rule{0.2cm}{0.08cm}}} Upper incisor with hard palate 
  }
    
    \label{fig:larynx_height}
\end{figure}

It is defined as the distance from the glottis to the hard palate right extremity (see~\autoref{fig:larynx_height}), used as the reference.

This measurement is particularly important for characterizing the vertical dimensions of the vocal tract, which have a direct influence on the resonance frequencies of the vocal tract.
    
    To perform this calculation, we first defined the center of the glottis as the geometric centroid of its contour points. The distance was then measured from this center to a specific reference point on the hard palate (see~\autoref{fig:larynx_height}), ensuring that this measurement was taken along an axis parallel to the pharyngeal wall.
This orientation guarantees anatomical relevance and provides better consistency with the vocal tract structures.
    
    To our knowledge, previous studies have never succeeded in measuring this specific variable of the vocal tract directly inside the vocal tract due to the limitations of electromagnetic articulography (EMA), which makes it impossible to position a sensor at the level of the larynx. The only option was to deduce the position of the larynx from video images of the speaker's head~\cite{Hoole1998}, which necessarily introduces a high degree of inaccuracy.
%PAV
\ifshowtodos
    \todo[inline]{PAV: Comme on peut considérer que c'est une première, ce serait bon de faire un petit paragraphe dans l'introduction pour dire que l'inversion ne s'est jamais attarder a sortir toutes ces variables articulatoire
    
    Et aussi d'ajouter que nous validons notre inversion a l'aune de ces variable articulatoire (directement dans le dernier paragramphe de l'intro). 
    }
\fi
%PAV
    Furthermore, on the MR imaging side several technical challenges have prevented this analysis: either the anatomical contours were not segmented accurately, or the anatomical structures were not sufficiently visible due to insufficient image resolution.
    
    Introducing this new vocal tract variable is a major advantage, since it provides a better understanding of the impact of vocal tract dimensions on the acoustic characteristics of speech, particularly on vocal resonances, or formants, which are fundamental for vowel perception and identification.
%PAV
\ifshowtodos
    \todo[inline]{PAV: Cest deux dernier paragraphe c'est plustôt a mettre dans l'introduction (c'est pas vraiment des méthodes ou de la stratégie de preuves). 
    }
\fi
%PAV

    \subsection{Calcul of TVs} 
    
    Unlike EMA data, which consist of discrete sensor points, our data comprise full articulatory contours. Before computing the tract variables, all contours were first denormalized. To estimate the relevant distances, we selected a specific range of points on each articulator and calculated the minimum Euclidean distance between these regions. This approach allows us to approximate constriction locations without relying on individual point measurements. Distances were computed for each tract variable (TV) and each frame independently, for both the original and predicted contours.

\iffalse
For each pair of points taken from the two articulators, we computed the Euclidean distance defined as:
    \begin{equation}
        d(\mathbf{x}, \mathbf{y}) = \sqrt{(x_1 - y_1)^2 + (x_2 - y_2)^2} 
        \label{eq3}
    \end{equation}
    
where \(\mathbf{x} = (x_1, x_2)\) and \(\mathbf{y} = (y_1, y_2)\) are two points in the plane.

The TV value is then defined as the smallest distance observed among all point pairs:
\begin{equation}
    \mathrm{TV} = \min_{\mathbf{x} \in A,\ \mathbf{y} \in B} d(\mathbf{x}, \mathbf{y})
    \label{eq4}
\end{equation}
where \(A\) and \(B\) denote the sets of points selected on the two articulators involved.
\fi

 The Pearson correlation, as shown in~(\autoref{eq:pearson}), was then used to quantify the relationship between each TV extracted from the original contours and those derived from the predicted contours, for all TVs.
    
    \begin{equation}
    \text{Pearson} = \frac{\sum_{i=1}^{n} (x_i - \bar{x}) (y_i - \bar{y})}
    {\sqrt{\sum_{i=1}^{n} (x_i - \bar{x})^2} , \sqrt{\sum_{i=1}^{n} (y_i - \bar{y})^2}}
    \label{eq:pearson}
    \end{equation}
    
    For velum movement, in addition to the Pearson correlation, we also used another metric. Since we consider any positive value as open and any negative value as closed, we represented these states as 1 and –1, respectively. To evaluate the agreement between the original and predicted TVs in this case, we used accuracy, as defined in~(\autoref{eq:accuracy}):
    
    \begin{equation}
    \text{Accuracy} = \frac{\text{Number of correct predictions}}{\text{Total number of predictions}}
    \label{eq:accuracy}
    \end{equation}
    
Accuracy provided better results than the Pearson correlation, which is why we decided to keep and report only the accuracy scores.

\subsection{Experiments}
Our primary objective is to successfully perform the inversion of the complete vocal tract contour under the best possible conditions. To this end, we conducted two distinct experiments.
The first experiment aimed to evaluate the impact of different acoustic input representations on the model’s performance. Three types of inputs were considered.
First, the Mel-Frequency Cepstral Coefficients (MFCCs), which are widely used in articulatory inversion studies and generally yield excellent results.
Second, LCCs, which provide higher-quality acoustic information. Indeed, some spectral peaks can be attenuated in MFCCs due to Mel-scale filtering and compression. LCCs, being more directly related to the spectral envelope of the vocal tract, enable the model —particularly in a single-speaker scenario— to establish a more direct mapping between acoustic features and articulatory trajectories.
Finally, we evaluated a self-supervised learning (SSL) embedding obtained from the HuBERT model, which has demonstrated promising results in previous studies \cite{attia2024improving}, even outperforming MFCC-based representations. Trained to predict masked acoustic representations, HuBERT produces rich and contextual embeddings capable of capturing fine-grained temporal and spectral information, making it particularly well-suited for complex tasks such as acoustic-to-articulatory inversion.

The second experiment focused on investigating the influence of dataset size on model performance. We examined whether similar results could be achieved using smaller subsets of datasets. To this end, the model was trained on datasets of increasing duration: the full 3.5-hour corpus, 2 hours, 1 hour, 30 minutes, and 10 minutes. In this experiment, the LCCs features were used as the model input and evaluations were conducted against the same complete test set.

\subsection{Model parameters}
    All models were trained for 300 epochs with a batch size of 10, using the Adam optimizer with an initial learning rate of 0.001. To avoid overfitting early stopping was applied with a patience of 10 epochs on the validation data, halting training if no improvement was observed. Our dataset contains 450,000 images after removing silence.We randomly divided our dataset by acquisitions into 80\% for training, 10\% for validation, and 10\% for testing.  
\iffalse
%PAV
\ifshowtodos
    \todo[inline]{PAV: Tu dois dire à quel niveau se fait le tigage alléatoire (session/séquence/sentence chunck). 
    En particulier le testset est il sur une session différente non vue ? 
    Et le tirage s'est fait une fois ou pour chaque durée de temps un tirage différent (10/30/60/120/210) et dans le cas ou c'est qu'une fois comment as tu réduit le data set (linéaire/alléatoire) ?}
\fi
%PAV
\fi

All experiments were run with the same training configuration and train-validation-test splits to ensure a fair comparison. The entire implementation was done using PyTorch.
\iffalse
%PAV
\ifshowtodos
    \todo[inline]{PAV: Il me semble que tu as oublier de parler de l'initialisation de ton model et si tu utilise la même initialisation pour chaque run de test ?}
\fi
%PAV
\fi

\subsection{Statistical Significance of the Results}
To evaluate whether the performance difference between different results is statistically significant, we accounted for the temporal dependency inherent in consecutive frames. The original test set contained 45,000 frames; however, consecutive frames are highly correlated. To ensure the independence of observations required for statistical testing, we aggregated the results by calculating the mean score for consecutive frames within the same phone. This process resulted in a reduced but more robust sample of $n = 10,871$ independent observations. Since the distribution is not normal (as tested with the D'Agostino test), the Wilcoxon Signed-Rank Test was used with a significance threshold of $\alpha = 0.05$.
%We first assessed the normality of the differences in performance using D'Agostino's $K^2$ test. The test rejected the null hypothesis of normality ($p < 0.05$). 
%Consequently, although the large sample size ($n > 10,000$) might allow for a Student's t-test via the Central Limit Theorem, we opted for the Wilcoxon Signed-Rank Test, a non-parametric alternative that does not assume a specific distribution.The Wilcoxon statistic $W$ is computed by ranking the absolute differences $|d_i|$ between the two models' scores:\begin{equation}W = \sum_{i=1}^{n} \text{rank}(|d_i|) \cdot \text{sgn}(d_i)\label{eq:wilcoxon}\end{equation}where $d_i = x_{1,i} - x_{2,i}$ represents the difference in performance for the $i$-th phoneme sequence. For large samples, the distribution of $W$ converges toward a normal distribution, allowing us to derive a Z-score and a p-value. We set the significance threshold at $\alpha = 0.05$.}

\begin{table*}[b]
\caption{Comparison of RMSE (mm) and median error (mm) for different input features: MFCCs, LCCs, and HuBERT embeddings.}
\label{tab:input_comparison}
\centering

% --- ZONE DE RÉGLAGES MANUELS ---
%\footnotesize      % OPTIONS : \scriptsize (petit), \footnotesize (moyen), \small (grand), \normalsize (normal)
\small
%\setlength{\tabcolsep}{10pt}    % Augmente ce chiffre pour élargir le tableau horizontalement
%\renewcommand{\arraystretch}{1.3} % Augmente ce chiffre pour agrandir la hauteur des lignes
% -------------------------------

\begin{tabular}{lcccccc} 
\toprule
 & \multicolumn{2}{c}{\textbf{MFCC-based model}} & \multicolumn{2}{c}{\textbf{HuBERT-based model}} & \multicolumn{2}{c}{\textbf{LCC-based model}} \\
\cmidrule(lr){2-3} \cmidrule(lr){4-5} \cmidrule(lr){6-7}
Articulator & RMSE & Median & RMSE & Median & RMSE & Median \\
\midrule
\textbf{Arytenoid}    & 1.63$^{*}$ $\pm$ 1.02 & 1.38 & 1.63$^{\phantom{*}}$ $\pm$ 1.00 & 1.39 & 1.60 $\pm$ 1.00 & 1.35 \\
\textbf{Epiglottis}   & 1.52$^{*}$ $\pm$ 0.88 & 1.33 & 1.49$^{*}$ $\pm$ 0.84 & 1.31 & 1.49 $\pm$ 0.85 & 1.31 \\
\textbf{Lower lip}    & 1.53$^{*}$ $\pm$ 0.83 & 1.35 & 1.48$^{\phantom{*}}$ $\pm$ 0.80 & 1.31 & 1.47 $\pm$ 0.81 & 1.30 \\
\textbf{Pharyngeal}   & 1.08$^{\phantom{*}}$ $\pm$ 0.58 & 0.96 & 1.09$^{\phantom{*}}$ $\pm$ 0.56 & 0.98 & 1.07 $\pm$ 0.57 & 0.94 \\
\textbf{Velum}        & 1.34$^{\phantom{*}}$ $\pm$ 0.67 & 1.21 & 1.35$^{\phantom{*}}$ $\pm$ 0.65 & 1.23 & 1.33 $\pm$ 0.66 & 1.20 \\
\textbf{Tongue}       & 2.33$^{*}$ $\pm$ 1.17 & 2.10 & 2.20$^{*}$ $\pm$ 1.12 & 1.99 & 2.28 $\pm$ 1.14 & 2.04 \\
\textbf{Upper lip}    & 1.15$^{*}$ $\pm$ 0.55 & 1.03 & 1.20$^{*}$ $\pm$ 0.58 & 1.09 & 1.11 $\pm$ 0.53 & 1.01 \\
\textbf{Vocal folds}  & 1.49$^{*}$ $\pm$ 0.84 & 1.31 & 1.47$^{*}$ $\pm$ 0.76 & 1.33 & 1.46 $\pm$ 0.80 & 1.30 \\ 
\midrule
\textbf{Mean}         & 1.51$^{*}$ $\pm$ 0.91 & 1.30 & 1.49$^{*}$ $\pm$ 0.87 & 1.31 & 1.48 $\pm$ 0.89 & 1.27 \\
\bottomrule
\end{tabular}

\vspace{6pt}
\begin{center}
  \scriptsize $^{*}$Significant difference compared to the LCC-based model result ($p < 0.05$) based on a Wilcoxon test.
\end{center}
\end{table*}

\section{Results}
First, we verified that our model does not overfit. For instance, the LCC-based model achieved an error of 1.48 mm on the test set, which is lower than the validation error of 1.63 mm. This slightly superior performance on the test set suggests that the model did not overfit.

All training runs converged in fewer than 150 epochs for both experiments, and all reported differences were tested for statistical significance using Wilcoxon test to ensure that the observed performance variations between embeddings and dataset sizes were meaningful.

The first experiment compared the model’s performance when trained with three different types of acoustic embeddings: MFCCs, LCCs, and HuBERT.
As shown in~\autoref{tab:input_comparison}, LCC-based model achieved the best overall performance across all articulators, with an average RMSE of 1.48 mm and a median error of 1.27 mm, followed by HuBERT-based model (RMSE = 1.49 mm, median = 1.31 mm) and MFCC-based model (RMSE = 1.51 mm, median = 1.30 mm). When examining each articulator individually, LCC-based model outperformed the other embeddings for six articulators and tied with HuBERT-based model for one — the epiglottis — where both reached an RMSE of 1.49 mm and a median of 1.31 mm. HuBERT-based model slightly surpassed LCC-based model for the tongue, with RMSE and median values of 2.20 mm and 1.99 mm, compared to 2.28 mm and 2.04 mm for LCC-based model. Overall, MFCC-based model consistently yielded the lowest performance across all articulators.

\begin{figure}[t]
    \centering
    \includegraphics[width=0.475\textwidth]{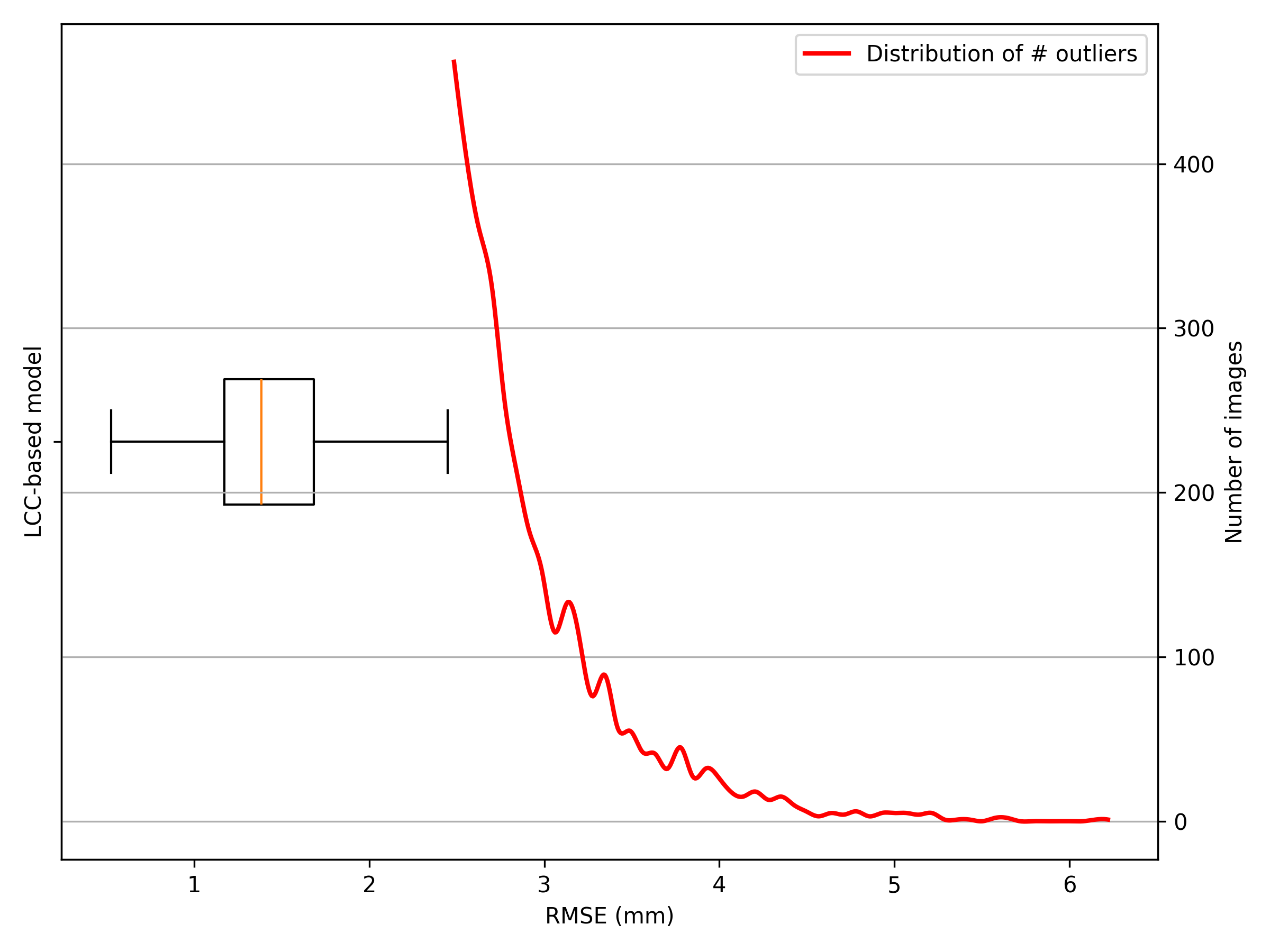}\\[2px]
    \caption{Boxplot of mean RMSE per frame for the LCC-based model, showing outlier ranges. The red curve shows the distribution of \# outliers. The right axis gives the number of outliers as a function of the error in mm.}
    \label{fig:boxplot}
\end{figure}

\autoref{fig:boxplot} shows a boxplot for the LCC-based model, illustrating the variability of the mean RMSE for each frame. It also indicates  the distribution of outliers.  

\begin{figure*}[b]
    \centering
    \includegraphics[width=1\textwidth]{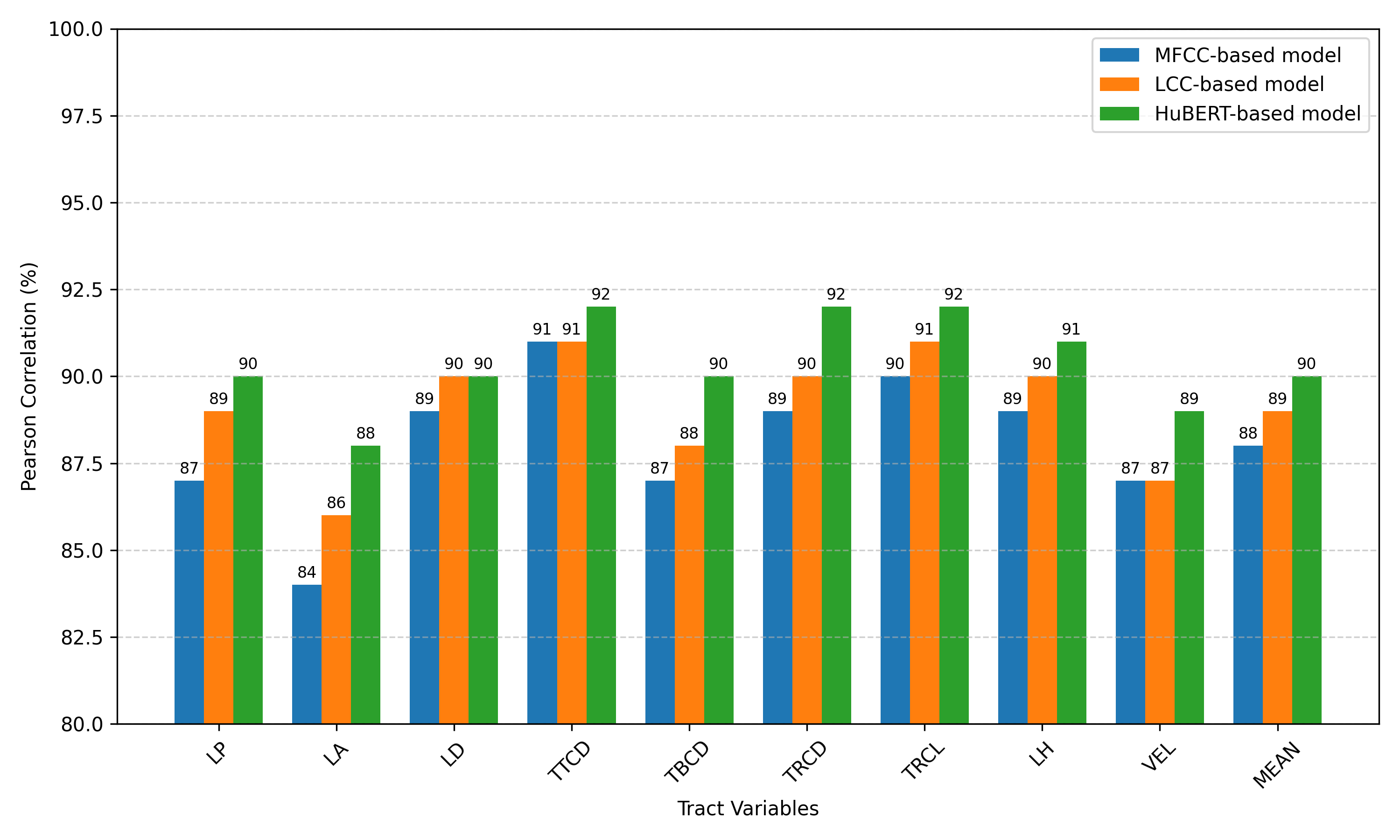}\\[2px]
    \caption{Correlation between the predicted and target trajectories for each measured tract variable.}
    \label{fig:VT_correlation}
\end{figure*}

\autoref{fig:VT_correlation} presents the Pearson correlations between ground-truth and predicted articulatory variables. HuBERT-based model achieved the highest correlations for most variables, except for LD, where it tied with LCC-based model (correlation = 0.91). Overall, HuBERT-based model’s correlations ranged from 0.88 to 0.92, followed by LCC-based model (0.86–0.91) and MFCC-based model (0.84–0.91).

\begin{figure*}[h]
    \centering
    \includegraphics[width=0.8\textwidth]{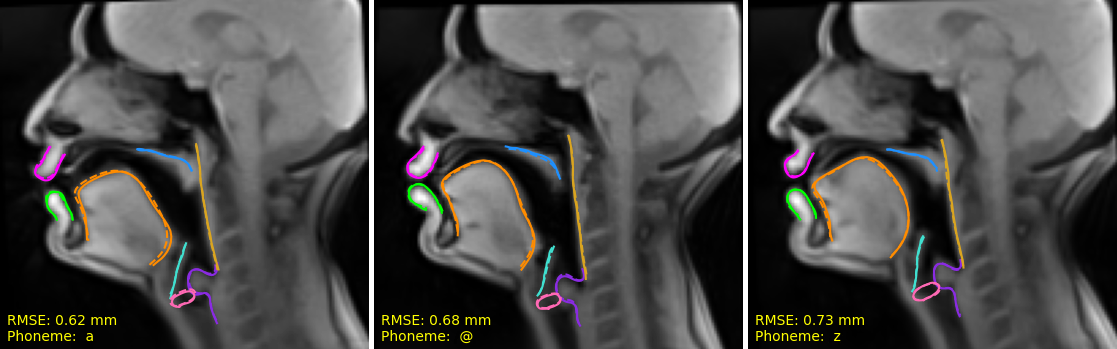}\\[2px]
    \includegraphics[width=0.8\textwidth]{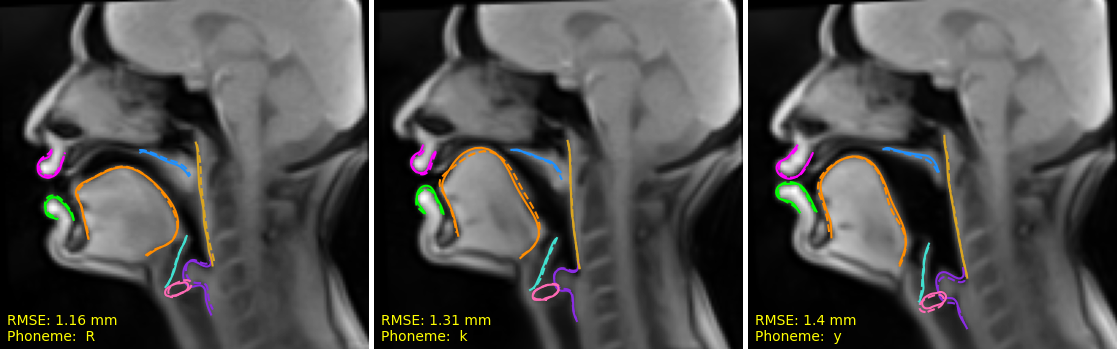}\\[2px]
    \includegraphics[width=0.8\textwidth]{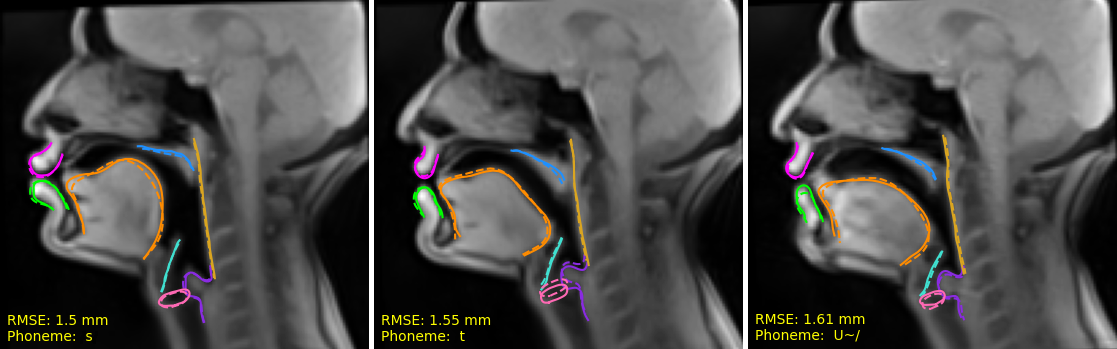}
    \caption{Illustration of inference images. The dotted lines represent the predicted values, while the solid lines correspond to the ground truth.}
    \label{fig:inference_images}
\end{figure*}

\autoref{fig:inference_images} shows examples of predicted and ground-truth contours using LCC-based model as input, with mean RMSE values ranging from 0.62 mm to 1.61 mm.

\begin{figure}[h]
    \centering
    \includegraphics[width=0.475\textwidth]{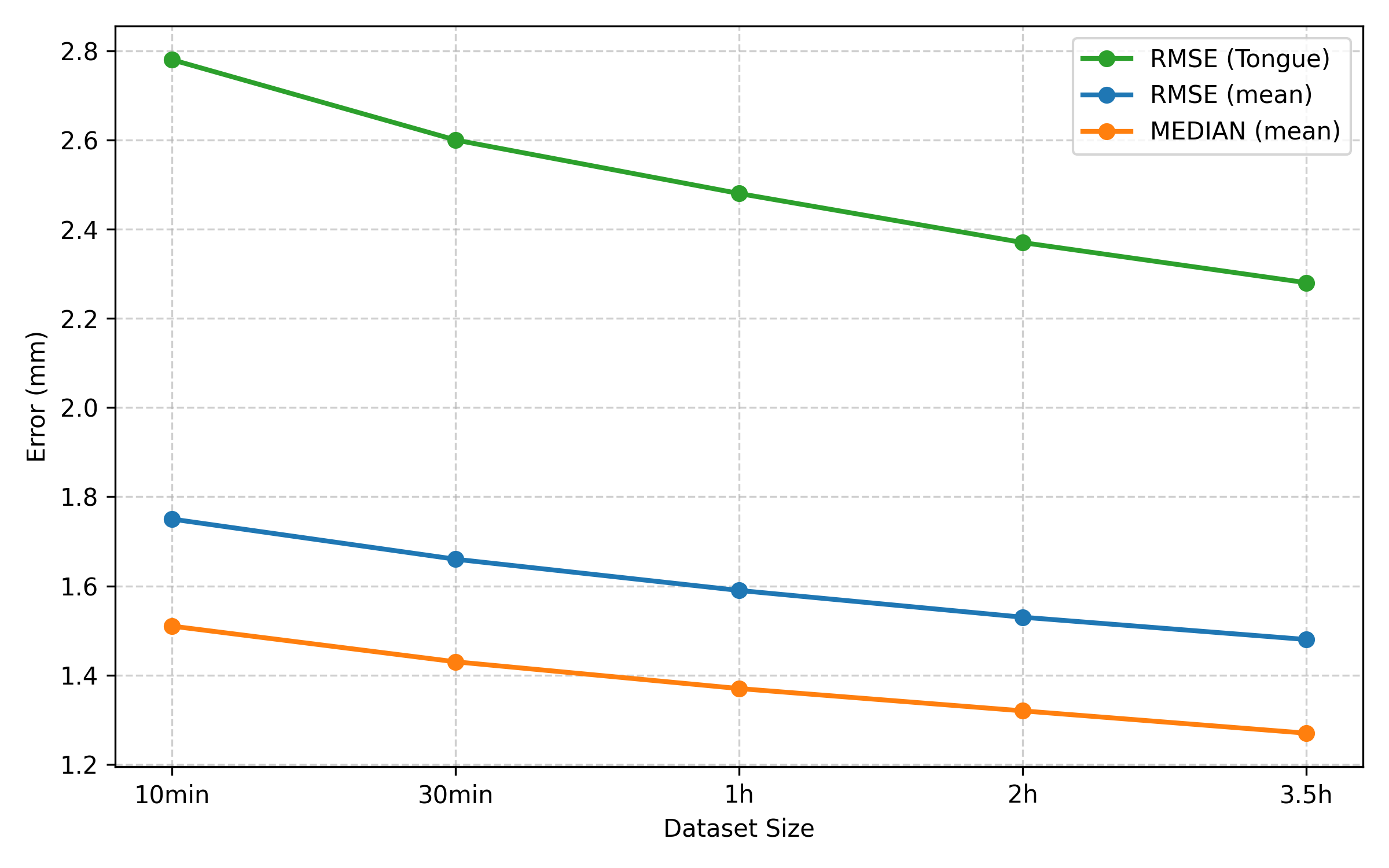}\\[2px]
    \caption{Comparison of tongue RMSE, mean RMSE, and median error by varying the dataset size.}
    \label{fig:size_comparison_graphe}
\end{figure}

\iffalse
\begin{table}[h]
\caption{Correlation between the predicted and target trajectories for each measured tract variable.}
\label{tab:VT_correlation}
\centering
\resizebox{0.7\linewidth}{!}{%
\begin{tabular}{|l|c|c|c|}
\hline
\multirow{2}{*}{\textbf{Tract variable}} & \multicolumn{3}{c|}{\textbf{Correlation}} \\
\cline{2-4}
                        & \textbf{MFCC-based model} & \textbf{LCC-based model} & \textbf{HuBERT-based model}\\
\hline
\textbf{LP}   & 0.87 & 0.89 & 0.90\\
\textbf{LA}   & 0.84 & 0.86 & 0.88\\
\textbf{LD}   & 0.89 & 0.90 & 0.90\\
\textbf{TTCD} & 0.91 & 0.91 & 0.92\\
\textbf{TBCD} & 0.87 & 0.88 & 0.90\\
\textbf{TRCD} & 0.89 & 0.90 & 0.92\\
\textbf{TRCL} & 0.90 & 0.91 & 0.92\\
\textbf{LH}   & 0.89 & 0.90 & 0.91\\
\textbf{VEL}  & 0.87 & 0.87 & 0.89\\
\textbf{VEL-PCA} & 0.83  & 0.84 & 0.87\\

\hline
\end{tabular}
}
\end{table}
\fi

%\caption*{\centering
%\small
%$^{*}$Significant difference compared to the ABA result ($p < 0.05$) based on a t-test.
%}

The second experiment evaluated the impact of dataset size on model performance using LCC-based model as input.
~\autoref{fig:size_comparison_graphe} shows the decrease in the mean RMSE from 1.75 mm to 1.48 mm, the mean median from 1.51 mm to 1.27 mm, as well as the tongue RMSE from 2.78 mm to 2.28 mm as the amount of training data increases from 10 minutes to 3.5 hours.

% A revoir 
Table~\ref{tab:size_comparison} further indicates that, when examining individual articulators, the 3.5-hour dataset produced the lowest RMSE and median values. Performance followed a clear hierarchical trend, with errors increasing as dataset size decreased from 2 hours to 1 hour, 30 minutes, and finally 10 minutes, which yielded the highest errors overall.

%: epiglottis (1.49 mm), lower lip (1.47 mm), velum (1.33 mm), tongue (2.28 mm), and vocal folds (1.46 mm). The arytenoid (1.58 mm) and pharyngeal region (1.06 mm) performed slightly better with the 1-hour dataset, while the upper lip achieved its best performance with the 2-hour and 30-minute subsets (1.10 mm).
% Regarding median errors, the 3.5-hour dataset provided the best results for six articulators and tied with the 1-hour and 2-hour datasets for the arytenoid (1.35 mm) and with the 2-hour dataset for the upper lip (1.01 mm). 

\begin{table*}[t]
\caption{Comparison of RMSE (mm) and MEDIAN (mm) by varying the size of the dataset.}
\label{tab:size_comparison}
\centering

% --- ZONE DE RÉGLAGES MANUELS ---
\footnotesize      % OPTIONS : \scriptsize (petit), \footnotesize (moyen), \small (grand), \normalsize (normal)

%\setlength{\tabcolsep}{10pt}    % Augmente ce chiffre pour élargir le tableau horizontalement
%\renewcommand{\arraystretch}{1.3} % Augmente ce chiffre pour agrandir la hauteur des lignes
% -------------------------------

\begin{tabular}{lcccccccccc}
\toprule
 & \multicolumn{2}{c}{\textbf{10 minutes}} & \multicolumn{2}{c}{\textbf{30 minutes}} & \multicolumn{2}{c}{\textbf{1 hour}} & \multicolumn{2}{c}{\textbf{2 hours}} & \multicolumn{2}{c}{\textbf{3.5 hours}}\\
\cmidrule(lr){2-3} \cmidrule(lr){4-5} \cmidrule(lr){6-7} \cmidrule(lr){8-9} \cmidrule(lr){10-11}
Articulator & RMSE & Median & RMSE & Median & RMSE & Median & RMSE & Median & RMSE & Median\\

\hline
\textbf{Arytenoid}           & 1,82$^{*}$ $\pm$\, 1.07 & 1,57 & 1.73$^{*}$ $\pm$\ 1.03 & 1.48 &1.70$^{*}$ $\pm$\ 1.01 & 1.45 & 1.65$^{*}$ $\pm$\ 1.01 & 1.41  & 1.60$\pm$\ 1.00 & 1.35 \\
\textbf{Epiglottis}          & 1.81$^{*}$ $\pm$\, 1.00 & 1.61 & 1.67$^{*}$ $\pm$\ 0.93 & 1.48 &1.61$^{*}$ $\pm$\ 0.92 & 1.41 & 1.57$^{*}$ $\pm$\ 0.88 & 1.38  & 1.49$\pm$\ 0.85 & 1.31 \\
\textbf{Lower lip}           & 1.83$^{*}$ $\pm$\, 0.99 & 1.61 & 1.78$^{*}$ $\pm$\ 0.97 & 1.57 &1.59$^{*}$ $\pm$\ 0.86 & 1.40 & 1.52$^{*}$ $\pm$\ 0.83 & 1.34  & 1.47$\pm$\ 0.81 & 1.30 \\
\textbf{pharyngeal}          & 1.22$^{*}$ $\pm$\, 0.66 & 1.08 & 1.15$^{*}$ $\pm$\ 0.61 & 1.03 &1.16$^{*}$ $\pm$\ 0.61 & 1.04 & 1.10$^{*}$ $\pm$\ 0.58 & 0.98  & 1.07$\pm$\ 0.57 & 0.94 \\
\textbf{Velum}               & 1.55$^{*}$ $\pm$\, 0.77 & 1,40 & 1.52$^{*}$ $\pm$\ 0.72 & 1.39 &1.43$^{*}$ $\pm$\ 0.71 & 1.29 & 1.37$^{*}$ $\pm$\ 0.68 & 1.24  & 1.33$\pm$\ 0.66 & 1.20 \\
\textbf{Tongue}              & 2.78$^{*}$ $\pm$\, 1.36 & 2.50 & 2.60$^{*}$ $\pm$\ 1.27 & 2.33 &2.48$^{*}$ $\pm$\ 1.24 & 2.22 & 2.37$^{*}$ $\pm$\ 1.20 & 2.11  & 2.28$\pm$\ 1.14 & 2.04 \\
\textbf{Upper lip}           & 1.27$^{*}$ $\pm$\, 0.60 & 1.15 & 1.21$^{*}$ $\pm$\ 0.57 & 1.10 &1.19$^{*}$ $\pm$\ 0.56 & 1.08 & 1.16$^{*}$ $\pm$\ 0.56 & 1.04  & 1.11$\pm$\ 0.53 & 1.01 \\
\textbf{Vocal folds}         & 1,72$^{*}$ $\pm$\, 0.95 & 1.52 & 1.62$^{*}$ $\pm$\ 0.88 & 1.43 &1.58$^{*}$ $\pm$\ 0.86 & 1.40 & 1.52$^{*}$ $\pm$\ 0.82 & 1.34  & 1.46$\pm$\ 0.80 & 1.30 \\ \midrule
\textbf{Mean}                & 1,75$^{*}$ $\pm$\, 1,05 & 1.51 & 1.66$^{*}$ $\pm$\ 0.99 & 1.43 &1.59$^{*}$ $\pm$\ 0.95 & 1.37 & 1.53$^{*}$ $\pm$\ 0.92 & 1.32  & 1.48$\pm$\ 0.89 & 1.27 \\
\bottomrule
\end{tabular}
\vspace{6pt}
\begin{center}
\scriptsize$^{*}$Significant difference compared to the 3.5 hours dataset result ($p < 0.05$) based on a Wilcoxon test.
\end{center}
\end{table*}

\section{Discussion}
\subsection{Comparative Analysis of Input Features}

~\autoref{tab:input_comparison} presents the results of the first experiment, which aims to compare different acoustic representations used as model inputs: MFCCs, LCCs, and HuBERT embeddings. The results show that LCC-based model provide the best performance, both globally (averaged over all articulators) and individually, except for the tongue, where HuBERT-based model achieves slightly better results. HuBERT-based model ranks second, while MFCC-based model yield the lowest performance. 
%\textcolor{soft-palate-midline}{However, it is important to note that all three embeddings exhibit errors exceeding the contour tracking error, which is approximately 1.00 mm.}

These results and their ranking can be explained by the specific nature of the information captured by each type of representation. LCCs are computed directly from the linear spectrum of the signal, without relying on the Mel scale, which is perceptually based. In the context of a single speaker, this spectral richness allows the model to learn a more direct mapping between acoustic features, especially spectral prominences corresponding to resonance frequencies, and articulatory trajectories. Indeed, for a given resonance cavity, the resonance frequencies are evenly spaced, so it is normal that LCCs corresponding to a linear frequency scale give better results. In other words, LCCs establish a more linear and physically interpretable relationship between the acoustic signal and the vocal tract geometry. However, this gain in robustness for a single speaker would disappear if multiple speakers had to be taken into account. 

HuBERT embeddings, derived from self-supervised learning, encode rich and contextual representations of the speech signal. By being trained to predict masked acoustic segments, HuBERT captures temporal dependencies and preserves fine acoustic variations that are directly related to articulatory configurations. Thus, although the results obtained with HuBERT-based model are highly satisfactory, they remain slightly below those of cepstral coefficients for the specific task of acoustic-to-articulatory inversion.

Finally, MFCCs—although widely used in speech processing—are based on a perceptual frequency scale (the Mel scale) and logarithmic compression. This design, oriented toward human auditory perception rather than the physical production of speech, attenuates high-frequency components that contain essential cues for accurately reconstructing articulatory movements. As a result, MFCC-based model show lower inversion accuracy compared to the other representations.

%\textcolor{soft-palate-midline}{To mitigate overfitting, we implemented an early stopping strategy with a patience of 10 epochs, performing validation checks every 10 epochs. For instance, the LCCs-based model achieved an error of 1.48 mm on the test set, which is lower than the validation error of 1.63 mm. This slightly superior performance on the test set suggests that the model has not overfitted and demonstrates strong generalization capabilities, with the validation set likely presenting a slightly higher degree of difficulty.}

According to~\autoref{fig:VT_correlation}, strong correlations are observed, exceeding 84\% for all tract variables. Among the three types of acoustic representations, HuBERT-based model consistently outperforms the others, achieving the highest correlations across all variables.
This superiority can be explained by the fact that, although it shows slightly higher RMSE and median values compared to LCC-based model, HuBERT-based model is able to accurately predict the contours of the articulators most involved in the production of the target phoneme.

However, it sometimes produces larger errors for articulators that are less involved in the same sound, as discussed in~\autoref{outlier}. This behavior is consistent with the nature of the model: HuBERT, based on self-supervised learning, extracts contextual and phonetically rich representations, while LCC-based model, grounded in a physical and deterministic modeling of the acoustic signal, maintain a more direct correspondence with articulatory reality.

Regarding MFCC-based model, although they show solid performance, their correlations remain slightly lower than those obtained with LCC-based model and HuBERT-based model. They effectively capture low-level acoustic properties through Mel-scale filtering and logarithmic compression, but their representational capacity remains limited for modeling articulatory dynamics and coarticulation effects.

%LCCs   : seuil = 3.73 mm Plus grande erreur : 6.25 mm
%HuBERT : seuil = 3.82 mm Plus grande erreur : 6.57 mm
%MFCCs  : seuil = 3.84 mm Plus grande erreur : 6,65 mm

It is not possible to compare our results with those in the literature because we have complete contours, whereas others only use information from a few points. However, regarding the correlation of articulatory variables, we obtain an overall correlation of 90\%, compared to 81\% in the case of~\cite{attia2024improving}.

\begin{table}[b]
    \caption{Number of outliers by type}
    \label{tab:outliers}
    \centering

    % --- ZONE DE RÉGLAGES MANUELS ---
    %\footnotesize      % Taille adaptée au format colonne
    \small
    %\setlength{\tabcolsep}{12pt}    % Espace entre les deux colonnes
    %\renewcommand{\arraystretch}{1.2} % Aération verticale
    % -------------------------------

    \begin{tabular}{lc} % 'l' pour le texte à gauche, 'c' pour le nombre au centre
        \toprule
        \textbf{Outlier type} & \textbf{\#} \\
        \midrule
        Image ambiguity              & 7  \\
        Global shift                 & 8  \\
        Tracking error               & 2  \\ 
        Critical articulator correct & 29 \\
        Inversion error              & 54 \\
        \bottomrule
    \end{tabular}
\end{table}

\subsection{Outliers} \label{outlier}
During our tests, we did not limit ourselves to evaluating the global average RMSE; we also examined the mean RMSE for each frame individually (see~\autoref{fig:boxplot}).

This analysis revealed that, for certain frames, the average RMSE reached very high values—exceeding 6 mm—compared to the global averages of 1.48 mm for LCC-based model
To identify these outliers, we computed the standard deviation (STD) and considered any value above the third quartile (Q3) as an outlier.

\begin{figure*}
    \centering
    \includegraphics[width=0.8\textwidth]{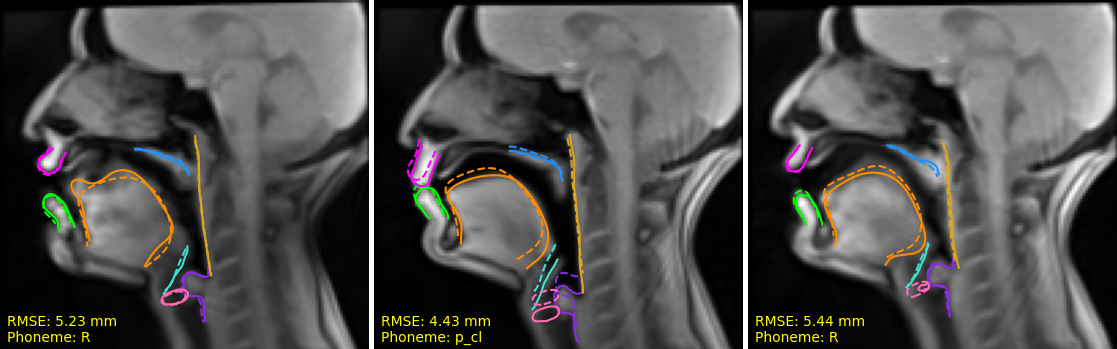}\\[2px]
    %\makebox[\textwidth]{%
        \includegraphics[width=0.535\textwidth]{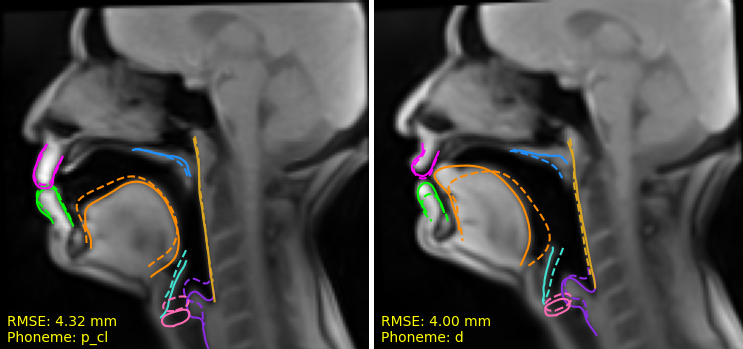}%
    %}\\
    \caption{Illustration of each of the 5 types of outliers. From top to bottom and  left to right: image ambiguity (tongue), global shift, tracking error (tongue), correct critical articulator (correct lip position), inversion error (tongue)}
    \label{fig:outlier}
\end{figure*}

We examined the 100 images with the most significant inversion outliers to better understand the origin of these errors. This investigation revealed five categories of outliers, summarized in ~\autoref{tab:outliers}, with the number of images corresponding to each category.~\autoref{fig:outlier} provides an illustration of each category.

The five categories are:\\
\begin{enumerate}
\item Images that are ambiguous due to MRI technical constraints during acquisition~\cite{Isaieva2021}. It is impossible to determine the contours and therefore to evaluate the inversion.
\item All articulator contours are shifted. The speaker's position has changed slightly, resulting in a slight translation of the position of all their articulators. The inversion therefore generates contours that are all shifted, leading to a significant outlier even though the overall shape of the vocal tract is correct.
\item The position of the critical articulator is correct, but the positions of the other articulators are incorrect. This concerns almost exclusively stops and the uvular consonant /\textinvscr/.
\item The contours produced by automatic tracking are incorrect, and the result of the inversion is closer to the actual shape of the vocal tract.
\item The inversion is incorrect, and one or more articulators are not in the expected position. 

%PAV
\ifshowtodos
    \todo[inline]{PAV: "This analysis revealed that .." toute cette partie jusqu'ici fait partie des methodes. }
\fi
%PAV

\end{enumerate}

This analysis shows that inversion errors account for just over half of the outliers. Categories 1 to 3 correspond to the process of forming the MRI images themselves, automatic contour tracking, or a small change in the speaker's posture. These three categories account for 17\% of the outliers. Category 4 is interesting because the critical articulator is correctly inverted. It accounts for 29\% of cases. 
%PAV
\ifshowtodos
    \todo[inline]{PAV: La partie ci dessus avec la table devrait être dans les résultats. }
\fi
%PAV
This mainly concerns plosives and, in some cases, nasal vowels. For /p/, for example (\autoref{fig:outlier}) (first image of the second row), the position of the lips is correct, but that of the other articulators is not. In this case, only the position of the lips is decisive, and the speaker can anticipate the position of the other articulators to minimize their articulatory effort. The last category is that of inversion errors. The position of one or more articulators is incorrect. For vowels, it is possible to achieve the same target in terms of formants with different shapes, even if this articulatory variability depends on the vowels \cite{potard:inria-00112226}, but we did not use acoustic simulation to verify that the inverted shapes could give the same acoustic parameters because, from a phonetic point of view, they seem very far from the expected result.
In summary, it is interesting to note that almost half of the outliers can be explained by problems other than the inversion process itself.

%PAV
\ifshowtodos
    \todo[inline]{PAV: Et la disscussion des autres résultats de la littérature en comparaison avec tes résultats, ressort les résultat de [30] et compare les explicitement en les citant (ce serait mieux si tu compare aussi avec Parrot [8]). C'est un minimum. Si tu trouve d'autre comparaison a faire avec la littérature c'est encore mieux, regarde dans les papiers que tu cite en introduction si tu peux pas y revenir pour comparer. la discussion ça sert pas seulement a discuter ses propre résultats mais a les comparer au reste de la littérature maintenant que le lecteur les connait.}
\fi
%PAV

\subsection{Impact of Dataset Size on Model Performance}
The results of the second experiment indicate that the 3.5-hour dataset produced the best overall performance for all articulators. Both the root mean square error (RMSE) and median error increase monotonically as the training duration decreases, with accuracy declining substantially for datasets smaller than 1 hour. Notably, the margin of improvement diminishes as the dataset grows; for instance, the difference in RMSE between the 10-minute and 30-minute subsets was $0.09$ mm, whereas the difference between the 3.5-hour and 2-hour datasets was only $0.05$ mm. This suggests that while increasing data volume consistently reduces error, performance has not yet stabilized, implying that even larger datasets may be required for full model convergence. Interestingly, the 1-hour subset achieved an average RMSE of $1.59$ mm—which is on the same order of magnitude as the pixel size ($1.62$ mm)—indicating that the model performs reasonably well even with a moderately sized dataset. This highlights the model's robustness and suggests that a 1-hour dataset may be sufficient for practical applications where data collection is costly or difficult to achieve.

\section{Conclusion}

%To the best of our knowledge, this is the first study that provides results for all the articulators of the vocal tract — from the glottis to the lips —  using RT-MRI images. Compared to our previous work~\cite{azzouz2025reconstruction}, we achieved better RMSE, i.e. 1.48\,mm instead of 1.65\,mm and a median error of 1.27\,mm instead of 1.47\,mm, relative to a pixel size of 1.62\,mm.

To our knowledge, this is the first study to achieve an RMSE of 1.48\,mm and a median error of 1.27\,mm across all vocal-tract articulators — from the glottis to the lips — using RT-MRI images. In comparison, our previous work~\cite{azzouz2025reconstruction} reported 1.65\,mm and 1.47\,mm, respectively. These findings demonstrate the feasibility of performing a complete vocal-tract inversion from RT-MRI data.

We also analyzed the articulatory variables for which HuBERT-based model achieved the highest correlation scores, all exceeding 88\%, demonstrating its strong capacity to model articulatory dynamics. In particular, the Larynx Height (LH) variable provides new insights into characterizing the vertical dimensions of the vocal tract, which have often been overlooked in prior studies.

%PAV
\ifshowtodos
    \todo[inline]{PAV: "accurate results" tu ne peut pas mettre une observation qualitative en conclusion. Donne quelque chose d'inconstestable : models with mean RMSE $<$ xxx and person corellation $>$ xxx on VT variables including new LH. Comme tu le dit après en fait enlève "accurate".}
\fi
%PAV

In terms of dataset size, to our knowledge, this is the largest RT-MRI corpus available for a single speaker, compared to those reported in the literature. This allowed us to conclude that our model trained with one hour of speech produced very satisfactory results, with only a very slight decrease in performance compared to three hours of data, i.e., 0.03\,mm in RMSE.

In addition, we evaluated the robustness of the model by testing it on speech recorded in a clean environment~\cite{azzouz2026acoustic}, following a training phase conducted exclusively on denoised speech. This study confirms the model's ability to adapt to changes in the acoustic characteristics of the input signal.
It is worth noting that the model was trained exclusively on a single speaker, which may limit its generalization to unseen speakers. Future work will focus on developing a speaker-independent model, where the potential of HuBERT-based model representations could be further demonstrated.

%\textcolor{red}{(On devrait supprimer ce paragraphe) Finally, training was carried out using a denoised audio signal of high quality. Although the denoising process preserved most of the relevant information, some subtle acoustic details may have been lost. Using clean, non-processed recordings could therefore lead to even better performance in future experiments.}

%PAV
\ifshowtodos
    \todo[inline]{PAV: "AAI" c'est la première occurence de l'accronyme tu dois le définir.}
\fi
%PAV
Beyond its theoretical significance in speech science, acoustics to articulatory inversion holds great potential for practical applications in speech synthesis, automatic speech recognition, clinical phonetics, and silent speech interfaces. This study aims to establish a robust and bidirectional link between articulatory gestures and their acoustic manifestations. Furthermore, it seeks to characterize the specific role of each articulator, providing insights that could help compensate for potential perturbations affecting individual articulatory mechanisms.

%PAV
\ifshowtodos
    \todo[inline]{PAV: Y a t'il un paragraphe sur code availlability qui est demandé par le journal ?}
\fi
%PAV

%PAV
\ifshowtodos
    \todo[inline]{PAV: L'entrée bibliographique 30 n'est pas complète.}
\fi
%PAV

%\clearpage
\bibliographystyle{IEEEtran}
\bibliography{ref}
\end{document}